\documentclass[10pt,showkeys,showpacs,aps,preprintnumbers]{revtex4}

\usepackage{graphicx}
\usepackage{amssymb}
\usepackage{bm}
\usepackage[sumlimits,intlimits]{amsmath}
\usepackage{amsfonts,amssymb}
\usepackage{amsmath}
\usepackage{bm}
\usepackage{graphics}
\usepackage{graphicx}
\usepackage{textcomp}
\usepackage{color}

\begin{document}


\title{Cosmologies with scalar fields from higher dimensions applied to Bianchi type $\rm VI_{h=-1}$ model:
classical and quantum solutions}
\author{J. Socorro}
\email{socorro@fisica.ugto.mx}
\author{L. Toledo Sesma}
\email{ltoledo@fisica.ugto.mx}
\author{Luis O. Pimentel$^2$}
\email{lopr@xanum.uam.mx}

\affiliation{$^1$Departamento de F\'{\i}sica, DCeI, Universidad de
Guanajuato-Campus Le\'on,
 C.P. 37150, Le\'on, Guanajuato, M\'exico\\
$^2$Departamento de F\'isica, Universidad Aut\'onoma Metropolitana,
Apartado Postal 55-534,C.P. 09340 M\'exico, DF, M\'exico}

\begin{abstract}
\vspace{1cm}
\begin{center}
{\bf Abstract}
\end{center}
In this work we construct an effective four-dimensional model by
compactifying a ten-dimensional theory of gravity coupled with a
real scalar dilaton field on a time-dependent torus. The
corresponding action in four dimensions is similar to the action of  K-essence
theories. This approach is applied to
 anisotropic cosmological Bianchi type $VI_{(h=-1)}$ model for which we study the classical coupling of the
 anisotropic scale factors with the two real scalar moduli produced by the compactification process.
 The classical Einstein field equations give us a hidden symmetry, corresponding to equal radii B=C, which permits us
to  solve exactly the equations of motion. With this hidden
symmetry, then we solve the FRW, finding that the scale factor goes
to B radii. Also the corresponding Wheeler-DeWitt (WDW) equation in
the context of Standard
 Quantum Cosmology is solved. Bohm's formalism for this cosmological model  is revisited  too.
\end{abstract}

\keywords{Exact solutions, classical and  quantum cosmology,
dimensional reduction} \pacs{98.80.Qc, 04.50.-h, 04.20.Jb, 04.50.Gh}
\date{\today}
\maketitle

\section{Introduction}

One of the most important things that we have learned from Planck's results \cite{Ade:2015bva} is related to the
little anisotropies of the universe. This result have played a tiny role in many theoretical results that describe
the dynamics of the universe, if we take the case of the Friedmann-Robertson-Walker (FRW) cosmological model we observe
that all the results in this road is developed in the cosmological principle scenario (homogeneous and isotropic universe).
 Another interesting aspect of cosmology is the inflationary process that the universe has  undergone in its early stages.
 In order to explain this process it has been necessary to introduce a scalar field in gravity theory that allows us
 to explain the accelerated expansion of the universe.
\\

The above problems have suggested to consider the presence of
higher-dimensional degrees of freedom in the cosmology derived from
four-dimensional effective theories. Some features of the presence
of higher-dimensional effective theories is to consider an effective
action with a graviton and a massless scalar field, the dilaton,
describing the evolution of the universe
\cite{Socorro:2015zfa,Sesma:2015zua}. On the other hand, it is well
known that relativistic theories of gravity, such as general
relativity or string theories, are invariant under reparametrization
of time. The quantization of such theories presents a number of
problems of principle known as the ``the problem of time''
\cite{kiefer2000conceptual,isham1993canonical}. This problem is
present in all systems, whose classical version is invariant under
time reparametrization, leading to its absence at the quantum
level. 
 Therefore, the formal
question involves how to handle the classical Hamiltonian
constraint, $\mathcal{H}\approx 0$, in the quantum theory. Also,
connected with the problem of time is the ``Hilbert space problem''
\cite{kiefer2000conceptual,isham1993canonical}
 referring to the not at all obvious selection of the inner product of states in quantum gravity, and whether there is a
 need for such a structure at all.
\\

In the present work we shall consider an alternative procedure about the role played by the moduli. In particular we shall not
 consider the presence of fluxes, as in string theory, in order to obtain a moduli-dependent scalar potential in the effective
 theory. Rather, we are going to promote some of the moduli to time-dependent by considering the particular case of a ten-dimensional
 gravity coupled to a time-dependent dilaton, compactified on a six-dimensional torus with a time-dependent K\"ahler modulus.
 With the purpose to track down the role play by such fields, we are going to ignore the dynamics of the complex structure field
 (for instance, by assuming that it is already stabilized by the presence of a string field in higher scales).
\\

It is very well known that the problem of time is present in all quantum cosmological models \cite{kiefer2000conceptual,isham1993canonical}.
There are some attempts to recover the notion of time for FRW models with matter given by a perfect fluid for
an arbitrary barotropic equation of state under the scheme of quantum cosmology (see \cite{Alvarenga:2001nm} for more details).

\section{Effective model}

We start from a ten-dimensional gravity theory coupled with a dilaton (dilaton field is the bosonic component common to all
superstring theories). In the string frame, the effective action depends on two space-time-dependent scalar fields:
the dilaton $\Phi(x^{\mu})$ and the K\"ahler modulus $\sigma(x^{\mu})$. For simplicity, in this work we shall assume
that these fields  depend  only on time. The higher dimensional (effective) theory is therefore given by

\begin{equation}
S=\frac{1}{2 \kappa^2_{10}} \int d^{10}X
\sqrt{-\hat{G}}\,e^{-2\Phi}\,\Bigg[\hat{\mathcal{R}}^{(10)}
+ 4\,\hat{G}^{MN}\nabla_{M}\,\Phi\,\nabla_{N}\,\Phi\Bigg] + \int
d^{10}X\,\sqrt{-\hat{G}}\,\mathcal{\hat L}_{m}
\label{action}
\end{equation}
where all quantities $\hat q$ refer to the string frame while the ten-dimensional metric is described by

\begin{equation}
ds^2=\hat{G}_{MN}\,dX^{M}\,dX^{N}= \hat{g}_{\mu\nu}\,dx^{\mu}dx^{\nu} + h_{mn}\,dy^{m}dy^{n},
\label{allgemeinemetric}
\end{equation}
where $M, N, P, \ldots$ are the indices of the ten-dimensional space, greek indices $\mu, \nu,\ldots =0,\ldots, 3$ and latin
indices $m, n, p,\ldots=4,\ldots,9$ correspond to the external and internal space, respectively. We will assume that the
six-dimensional internal space has the form of a torus with a metric  given by

\begin{equation}
h_{mn}=e^{-2\sigma(t)}\,\delta_{mn},
\label{intmet}
\end{equation}
with $\sigma$ a real parameter. By reducing the higher dimensional action \eqref{action} to four dimensions and rewrite
it in the Einstein frame (for more details see \cite{Socorro:2015zfa,Sesma:2015zua}) this gives the reduced action

\begin{equation}
S_4=\frac{1}{2\kappa^2_{4}}\int d^4x \sqrt{-g}\,\biggl(\mathcal{R}
-2\,g^{\mu\nu}\,\nabla_{\mu}\phi\nabla_{\nu}\phi -
96\,g^{\mu\nu}\,\nabla_{\mu}\sigma\nabla_{\nu}\sigma -
36\,g^{\mu\nu}\,\nabla_{\mu}\phi\nabla_{\nu}\sigma + \mathcal{L}^{(4)}_{\text{matter}} \biggr),
\label{endeaction}
\end{equation}
where $\phi=\Phi+\frac{1}{2}ln(\hat{V})$, with $\hat{V}$ given by

\begin{equation}
\hat{V}=e^{6\sigma(t)}Vol(X_6)=\int d^6y,
\end{equation}
being the volume associated to the six-dimensional space. By
considering only a time-dependence on the moduli, one can notice
that for the internal volume $Vol(X_6)$ to be small, the modulus
$\sigma(t)$ should be a monotonic increasing function on time
(recall that $\sigma$ is a real parameter), while $\hat{V}$ is time
and moduli independent. Recently it has appeared an article related
with our ideas. They work in the context of quantum cosmological
models in a $n$-dimensional anisotropic space by introducing
massless scalar fields \cite{Alves-Junior:2016vpw}. It will be our
subject as it was in our previous results
\cite{Socorro:2015zfa,Sesma:2015zua}

\ignorespaces

We can mention that the moduli fields will satisfy a Klein-Gordon like equation in the Einstein frame as an effective
theory. This will be possible to appreciate  by taking  a variation of the action \eqref{endeaction} with respect to each of
the moduli fields. Now with the expression \eqref{endeaction} we proceed to build up the Lagrangian and the Hamiltonian of
the theory at the classical regime employing the anisotropic cosmological Bianchi type $VI_{h=-1}$ model. This is the subject of
the subsection \ref{EEE}.

\subsection{Effective Einstein equations in four dimensions}\label{EEE}

The equation of motions associated to the reduced action in four dimensions \eqref{endeaction} can be obtained by taking
variations with respect to each of the fields. So, in this sense we have that the Einstein equations and the Klein-Gordon
like equations (EKG) associated to the fields $(\phi, \sigma)$ are given by

\begin{subequations}
\begin{eqnarray}
R_{\alpha\beta} - \frac{1}{2}\,g_{\alpha \beta}\,R & = & 2\left(\nabla_{\alpha}\phi\,\nabla_{\beta}\phi
- \frac{1}{2}\,g_{\alpha\beta} \nabla^{\gamma}\,\phi\nabla_{\gamma}\phi\right)
+ 96\,\left(\nabla_{\alpha}\sigma\,\nabla_{\beta}\sigma - \frac{1}{2}\,g_{\alpha\beta}\,\nabla^{\gamma}\sigma\,\nabla_{\gamma}\sigma
\right)\nonumber\\
& & +\, 36\,\left(\nabla_{\alpha}\phi\,\nabla_{\beta}\sigma - \frac{1}{2}\,g_{\alpha\beta}\,\nabla^{\gamma}\phi\,\nabla_{\gamma}\sigma
\right) + 8\pi G\,T_{\alpha\beta},
\label{einstein}\\
\Box\,\phi & = & 0, \label{phi}\\
\Box\,\sigma & = & 0, \label{sigma}
\end{eqnarray}
\label{EKG}
\end{subequations}
where $\Box$ is the d'Alembertian operator in four dimensions, which is written as $\Box=g^{\alpha\beta}\partial_{\alpha}\partial_{\beta}$.
  Since we are interested in anisotropic background, we are going to assume that the four dimensional metric $g_{\alpha\beta}$ is
  described by the Bianchi type $\rm VI_{h=-1}$ model which line element can be read
  as (we write in usual way and in the Misner's parametrization)

\begin{eqnarray}
\rm ds^2&=&\rm -N^2 dt^2 + A^2(t)\, dx^2 + B^2(t) e^{-2x}\,dy^2+
C^2(t)e^{2x}\, dz^2, \label{metric-vi} \\
&& -N^2 dt^2 + e^{2\Omega-4\beta_+}\, dx^2 + e^{2\Omega+2\beta_++
2\sqrt{3} \beta_- -2x}\,dy^2+ e^{2\Omega+2\beta_+- 2\sqrt{3}
\beta_-+2x}\, dz^2, \label{misner-p}
\end{eqnarray}
where $\rm N(t)$ is the lapse function, the functions $\rm A(t)$,
$\rm B(t)$ and $\rm C(t)$ are the corresponding scale factors in the
directions $\rm (x,y,z)$, respectively, also using the Misner's
parametrization for the radii in this model,
$$\rm A=e^{\Omega-2\beta_+}, \qquad B=e^{\Omega+\beta_++ \sqrt{3}
\beta_-}, \qquad C=e^{\Omega+\beta_+- \sqrt{3} \beta_-}.$$

Writing the field equations \eqref{EKG} in the background metric
\eqref{metric-vi}, we see that the equation of motions
\eqref{einstein}
are given by
\begin{subequations}
\begin{eqnarray}
\rm G^0_0: && \rm  2\,\frac{A'}{A}\frac{C'}{C} +
\left(\frac{C'}{C}\right)^2 - \phi'^2 - 18 \phi'\sigma' -
48\sigma'^2 - 8\pi G \rho
- \frac{1}{A^2}=0, \label{00new} \\
\rm G^{0}_{1} = -\,G^{1}_{0} : &&  \rm \frac{B'}{B} - \frac{C'}{C} = 0, \label{01new}\\
\rm G^1_1: &&  \rm 2\,\frac{C''}{C} + \left(\frac{C'}{C}\right)^2 +
\phi'^2 + 18\phi'\sigma' + 48\sigma'^2 + 8\pi G P
+ \frac{1}{A^2}=0,\label{11new}\\
\rm G^2_2: && \rm \frac{A''}{A} + \frac{A'}{A}\frac{C'}{C} +
\frac{C''}{C} + \phi'^2 + 18\phi'\sigma' + 48\sigma'^2 + 8\pi G P
- \frac{1}{A^2}=0,\label{22new}\\
\rm \Box\phi=0: && \rm 2\,\left(\frac{A'}{A} + \frac{C'}{C}\right)\phi' + \phi''=0, \label{dalaphi}\\
\rm \Box\sigma=0: && \rm 2\,\left(\frac{A'}{A} +
\frac{C'}{C}\right)\sigma' + \sigma''=0, \label{dalasigma}
\end{eqnarray}
\end{subequations}
the other components can be seen in the appendix \ref{appe}, and
$(\ldots)'$ means derivative with respect to the proper time $\rm
d\tau=N(t)\,dt$. 
 From the last expressions,
we can see that it is easy to solve the equation \eqref{01new},
whose solutions is $\rm B(t)=\rm C(t)$ (see the expression
\eqref{01} of the appendix \ref{appe} for more detailed derivation).
This tell us that two scale factors evolve in the same way. From
this result, we see also that the components $G_{2}^2 = G_{3}^3$ (as
we can observe from the expressions \eqref{22}, and \eqref{33}). On
the other hand, the solution for the fields $(\phi,\sigma)$ are
obtained from equations \eqref{dalaphi}, and \eqref{dalasigma} whose
solutions can be read as

\begin{equation}
\phi(\tau)=\phi_0\int \frac{d\tau}{AC^2} + \phi_1, \qquad\qquad
\sigma(\tau)=\sigma_0\int \frac{d\tau}{AC^2} + \sigma_1.
\end{equation}
By taking the relation between the proper and cosmic time ($\rm
d\tau=N(t)dt$) we see that the lapse function $\rm N(t)$ can be
choose as $\rm N\sim AC^2$. This tell us that the lapse function
plays the role of a gauge transformation. So, under this scheme it  is
possible to obtain a  simpler solution in the cosmic time t, as
\begin{equation}
\phi(\tau)=\phi_0 t + \phi_1,\qquad\qquad \sigma(\tau)=\sigma_0 t + \sigma_1,
\label{phisig}
\end{equation}
where $(\phi_0,\phi_1,\sigma_0,\sigma_1)$ are integration constants.
\\

So far, we have found from the fields equations \eqref{EKG} that the
radii B and C are equal for the Bianchi $VI_{h=-1}$, this classical
hidden symmetry is relevant in the quantum level, because 21 years
ago, was found a generic quantum solutions for all Bianchi Class A
cosmological models in the Bohm's formalism
\cite{socorro-obregon:1996}, and in particular for the Bianchi type
$\rm VI_{h=-1}$ it  was necessary to modify the general structure of the
generic solution with a function over the coordinate
$\beta_1-\beta_2$, with this result, the modification is not
necessary, due  to the fact that this function is a constant now, as we see using
the Misner's parametrization of this cosmological model. On the
other hand, the moduli fields $(\phi,\sigma)$ have the linear
behavior in  time, as we can see in the expressions
\eqref{phisig}.
 These solutions were  obtained by taking the gauge $N\sim AC^2$.  This gauge choice will play a great role in the
 next section, when we deal with the classical Lagrangian and Hamiltonian which we shall obtain in the next section.

\subsection{Misner parametrization}
Using the Misner parametrization for the radii in this model,
$$\rm A=e^{\Omega-2\beta_+}, \qquad B=e^{\Omega+\beta_++ \sqrt{3}
\beta_-}, \qquad C=e^{\Omega+\beta_+- \sqrt{3} \beta_-},$$ with
this, The EKG classical equations
(\ref{einstein},\ref{phi},\ref{sigma}), using this parametrization
become

\begin{eqnarray}
\rm G^0_0:&& 3\frac{\dot \Omega^2}{N^2}-3\frac{\dot
\beta_+^2}{N^2}-3\frac{\dot \beta_-^2}{N^2}-\frac{\dot
\phi^2}{N^2}-18\frac{\dot \phi \dot \sigma}{N^2}-48 \frac{\dot
\sigma^2}{N^2}-8\pi G \rho - e^{-2\Omega +4\beta_+}=0, \label{00-m} \\
  \rm G^1_1:&& \rm 2\frac{\ddot
\Omega}{N^2}+3\frac{\dot \Omega^2}{N^2}+ 6\frac{\dot \Omega \dot
\beta_+}{N^2}-2\frac{\dot \Omega \dot N}{N^3} +2\frac{\ddot
\beta_+}{N^2} +3\frac{\dot \beta_+^2}{N^2}-2\frac{\dot \beta_+ \dot
N}{N^3}+3\frac{\dot
\beta_-^2}{N^2} \nonumber\\
&&\rm \quad +\frac{\dot \phi}{N^2}+18\frac{\dot \phi \dot
\sigma}{N^2}+48\frac{\dot \sigma^2}{N^2}+8\pi G P + e^{-2\Omega
+4\beta_+}=0,\label{11-m}\\
\rm G^2_2:&& \rm 2\frac{\ddot \Omega}{N^2}+3\frac{\dot
\Omega^2}{N^2}- 3\frac{\dot \Omega \dot \beta_+}{N^2}-2\frac{\dot
\Omega \dot N}{N^3} -3\sqrt{3} \frac{\dot \Omega \dot
\beta_-}{N^2}-\frac{\ddot \beta_+}{N^2} +3\frac{\dot
\beta_+^2}{N^2}+\frac{\dot \beta_+ \dot N}{N^3}-\sqrt{3} \frac{\ddot
\beta_-}{N^2}+3\frac{\dot
\beta_-^2}{N^2} +\sqrt{3}\frac{\dot \beta_- \dot N}{N^3} \nonumber\\
&&\rm \quad +\frac{\dot \phi}{N^2}+18\frac{\dot \phi \dot
\sigma}{N^2}+48\frac{\dot \sigma^2}{N^2}+8\pi G P - e^{-2\Omega
+4\beta_+}=0,\label{22-m}\\
\rm G^3_3:&& \rm 2\frac{\ddot \Omega}{N^2}+3\frac{\dot
\Omega^2}{N^2}- 3\frac{\dot \Omega \dot \beta_+}{N^2}-2\frac{\dot
\Omega \dot N}{N^3} +3\sqrt{3} \frac{\dot \Omega \dot
\beta_-}{N^2}-\frac{\ddot \beta_+}{N^2} +3\frac{\dot
\beta_+^2}{N^2}+\frac{\dot \beta_+ \dot N}{N^3}+\sqrt{3} \frac{\ddot
\beta_-}{N^2}+3\frac{\dot
\beta_-^2}{N^2} -\sqrt{3}\frac{\dot \beta_- \dot N}{N^3} \nonumber\\
&&\rm \quad +\frac{\dot \phi}{N^2}+18\frac{\dot \phi \dot
\sigma}{N^2}+48\frac{\dot \sigma^2}{N^2}+8\pi G P - e^{-2\Omega
+4\beta_+}=0,\label{33-m}\\
\rm G_{01}=G_{10}:&& \rm 2\sqrt{3} \dot \beta_-=0, \label{01-m}\\
 \rm \Box \phi=0:&& -3\frac{\dot \Omega}{N}\frac{\dot \phi}{N}- \frac{\ddot
\phi}{N^2}+ \frac{\dot \phi}{N}\frac{\dot N}{N^2}=0,
\label{dalaphi0-m}\\
\rm \Box \sigma=0:&& -3\frac{\dot \Omega}{N}\frac{\dot \sigma}{N}-
\frac{\ddot \sigma}{N^2}+ \frac{\dot \sigma}{N}\frac{\dot N}{N^2}=0,
\label{dalasigma0-m}
\end{eqnarray}
equation (\ref{01-m}) imply that $\rm \beta_-=\beta_0=constant$,
then the last set of equations is read as (using the time
parametrization $\rm d\tau =N dt$, $\prime=\frac{d}{d\tau}$)
\begin{eqnarray}
\rm G^0_0:&& \rm 3 \Omega^{\prime 2}-3 \beta_+^{\prime 2}-
\phi^{\prime 2}-18\phi^{\prime}  \sigma^{\prime}-48
\sigma^{\prime 2}-8\pi G \rho - e^{-2\Omega +4\beta_+}=0, \label{00n} \\
  \rm G^1_1:&& \rm 2
\Omega^{\prime \prime}+3 \Omega^{\prime 2}+ 6\Omega^{\prime}
\beta_+^{\prime} +2 \beta_+^{\prime \prime} +3\beta_+^{\prime 2} +
\phi^{\prime}+18 \phi^{\prime} \sigma^{\prime}+48 \sigma^{\prime
2}+8\pi G P + e^{-2\Omega
+4\beta_+}=0,\label{11n}\\
\rm G^2_2:&& \rm 2 \Omega^{\prime \prime}+3 \Omega^{\prime 2}- 3
\Omega^{\prime} \beta_+^{\prime} - \beta_+^{\prime \prime } +3
\beta_+^{\prime 2} + \phi^{\prime}+18 \phi^{\prime}
\sigma^{\prime}+48 \sigma^{\prime 2}+8\pi G P - e^{-2\Omega
+4\beta_+}=0,\label{22n}\\
 \rm && 3 \Omega^{\prime} \phi^{\prime}+
\phi^{\prime \prime}=0,
\label{dalaphi0-mm}\\
 \rm && 3 \Omega^{\prime} \sigma^{\prime}+
\sigma^{\prime \prime}=0,  \label{dalasigma0-mm}
\end{eqnarray}
The solutions for the fields $(\phi,\sigma)$ are obtained from
equations (\ref{dalaphi0-mm},\ref{dalasigma0-mm}) as
\begin{equation}
\rm \phi(\tau)=\phi_0\int e^{-3\Omega} d\tau + \phi_1,\qquad\qquad
\sigma(\tau)=\sigma_0\int e^{-3\Omega} d\tau +\sigma_1,
\end{equation}
that in the gauge $\rm N  \thicksim e^{3\Omega}$, we have the
simplest solution in the cosmic time t, as
\begin{equation}
\rm \phi(\tau)=\phi_0 t + \phi_1,\qquad\qquad \sigma(\tau)=\sigma_0
t +\sigma_1,
\end{equation}
 where $(\phi_0,\phi_1,\sigma_0,\sigma_1)$
are integration constants, which are the same solution as in the
previous parametrization of the metric.

\section{Classical Lagrangian and Hamiltonian}\label{CLH}

We have found in the last section that the Bianchi $VI_{h=-1}$ has two equal radii.  With the idea to reach
the quantum regime of our model, we shall develop the classical Lagrangian and Hamiltonian analysis.
Through this picture we will find that we have one conserved
quantity and this give us a first constraint in the Hamiltonian formulation. We start by looking for the Lagrangian
density for the matter content. This is given by a barotropic perfect fluid, whose stress-energy tensor
is \cite{Socorro:2003fn,Chowdhury:2006pk}

\begin{equation}
T_{\mu\nu}=(p+\rho) u_\mu u_\nu + p g_{\mu \nu},
\label{stress}
\end{equation}
which satisfies the conservation law $\nabla_{\nu}T^{\mu\nu}=0$. Taking the equation of state $p=\gamma\rho$ between the energy density
$\rho$ and the pressure $p$ of the comoving fluid, we see that a solution is given by $\rho=C_{\gamma}(AC^2)^{-(1+\gamma)}$ with
$C_{\gamma}$ the corresponding constant for different values of $\gamma$ related to the universe evolution stage. Then, the Lagrangian
density for the matter content reads

\begin{equation}
\mathcal{L}_{\text{matter}}\,=\,16\pi\,G_{N}\,\sqrt{-g}\,\rho\,=\,16\pi\,NG_{N}\rho_{\gamma}\,(AC^2)^{-(1+\gamma)}.
\label{mattcont}
\end{equation}
And then the Lagrangian that describes the fields dynamics is given by

\begin{equation}
\mathcal{L}_{VI}= 2\frac{A\dot{C^2}}{N} + 4\frac{C\dot{A}\dot{C}}{N} + 2\frac{AC^2\dot{\phi^2}}{N}
+ 36\frac{AC^2\dot{\phi}\dot{\sigma}}{N} + 96\frac{AC^2\dot{\sigma^2}}{N} + 2\frac{C^2}{A}N + 16\pi GAC^2\rho N.
\label{lagra-vi}
\end{equation}
In the last expression we have employed the result that the radii $B$ and $C$ are equal. Using the standard definition of the momenta
$\Pi_{q^{\mu}}=\frac{\partial\mathcal{L}}{\partial\dot{q}^{\mu}}$, where $q^{\mu}$ are the coordinate fields $q^{\mu}=(A,C,\phi,\sigma)$
we obtain the momenta associated to each field

\begin{alignat}{2}
\Pi_{A} &= \frac{4C\dot C}{N}, &\qquad   \dot{C} &= \frac{N\Pi_{A}}{4C}, \nonumber\\
\Pi_{C} &= 4\frac{A\dot C}{N} + 4\frac{C\dot A}{N}, &\qquad  \dot{A} &= \frac{N}{4AC^2}\left[C\Pi_{C} - A\Pi_{A}\right],\nonumber\\
\Pi_{\phi} &= \frac{AC^2}{N}\left[36\dot{\sigma} + 4\dot{\phi}\right], &\qquad \dot{\phi} &= \frac{N}{44AC^2}\left[3\Pi_{\sigma}
- 16\Pi_{\phi}\right], \nonumber\\
\Pi_{\sigma} &= \frac{AC^2}{N}\left[192\dot{\sigma} + 36\dot{\phi}\right], &\qquad \dot{\sigma} &= \frac{N}{132AC^2}
\left[9\Pi_{\phi} - \Pi_{\sigma}\right], \nonumber
\end{alignat}
and introducing them into the Lagrangian density, we obtain the canonical Lagrangian as
$\mathcal{L}_{canonical} =\Pi_{q^\mu} \dot q^\mu - N\mathcal{H}$. When we perform the variation of this canonical
Lagrangian with respect to $N$,  $\frac{\delta\mathcal{L}_{canonical}}{\delta N} =0$, we obtain the constraint $\mathcal{H}=0$.
In our model the only constraint corresponds to Hamiltonian density, which is weakly zero. So, we obtain the Hamiltonian density for this model

\begin{equation}
\mathcal{H}_{VI}= \frac{1}{8AC^2}\left[-A^2\Pi_A^2 +2A\Pi_A \,C\Pi_C - \frac{16}{11}\Pi_\phi^2 + \frac{6}{11}\Pi_\phi\,\Pi_\sigma
- \frac{1}{33}\Pi_\sigma^2 - 128\,\pi\,G_{N}\,\rho_{\gamma} \left(AC^2\right)^{1-\gamma} - 16 C^4\right].
\label{ham}
\end{equation}
The last Hamiltonian can be rewritten in a simpler  way by taking the transformations related with the generalized coordinates and momenta
 as $Q=e^q$, and $\Pi_{Q}=\frac{\partial S}{\partial Q}$. So we see that the momenta can be transformed
 as follow $\Pi_A\to\Pi_{A}=e^{-a}P_{a}$, where $P_{a}=\frac{\partial S}{\partial
 a}$, and we write in similar way the other momenta in the field $\rm (\phi,\sigma)$, is say
  $\rm \Pi_\phi=P_\phi$, $\rm \Pi_\sigma=P_\sigma$.
  Now, we develop our analysis for the case when the matter content is taken as a stiff fluid, $\gamma=1$.
  In this case the Hamiltonian density takes the following form

\begin{equation}
\mathcal{H}=-P_{a}^2 + 2P_{a} P_{c} - \frac{16}{11}P_{\phi}^2 +
\frac{6}{11}P_{\phi}\,P_{\sigma} - \frac{1}{33}P_{\sigma}^2 -
16e^{4c} - 128\,\pi\,G_{N}\,\rho_{1}, \label{ham-ac}
\end{equation}
where we have used the gauge transformation $N=8AC^2$. By using the Hamilton equations for the momenta
 $\dot{P}_{\mu}=-\frac{\partial\mathcal{H}}{\partial q^\mu}$ and coordinates $\dot{q}^\mu =\frac{\partial\mathcal{H}}{\partial P_{\mu}}$,
  we have

\begin{subequations}
\begin{eqnarray}
\dot{a}&=& -2P_{a} + 2P_{c}, \label{a}\\
\dot{c}&=& 2P_{a}, \label{c}\\
\rm \dot \phi &=& \rm -\frac{32}{11}P_\phi+
\frac{6}{11}P_\sigma, \label{f-phi}\\
\rm \dot \sigma &=& \rm \frac{6}{11}P_\phi-\frac{2}{33}P_\sigma, \label{f-sigma}\\
\rm \dot{P}_a&=& \rm  0, \qquad \qquad \qquad \Rightarrow \qquad P_a= p_{a}=constant, \label{pa}\\
\rm \dot{P}_c&=& \rm 64 e^{4c}, \label{pc}\\
\rm \dot{P}_{\phi}&=& \rm 0, \qquad \qquad \qquad \Rightarrow \qquad P_\phi=p_{\phi}=constant, \label{pphi}\\
\rm \dot{P}_{\sigma}&=& \rm 0, \qquad \qquad \qquad \Rightarrow
\qquad P_\sigma=p_{\sigma}=constant.
\end{eqnarray}
\label{psigma}
\end{subequations}
From the last system of differential equations it is possible to
solve the equations \eqref{c} by taking the solution associated to
the momenta $\rm P_{a}$ which is constant. we have $\rm
c(t)=2p_{a}\,t + c_{o}$, and by transforming this solution to the
original generalized coordinate, we obtain $\rm
C(t)=C_{0}e^{2p_{a}t}$. On the other hand, to solve the momenta $\rm
P_{c}$ we use the Hamiltonian density to obtain $\rm 16
e^{4c}=2P_{a} P_{c} + \alpha$, where $\rm \alpha = -p_{a}^2 -
\frac{16}{11}p_{\phi}^2 + \frac{6}{11}p_{\phi}\,p_{\sigma} -
\frac{1}{33} p_{\sigma}^2- 128\,\pi\,G_{N}\,\rho_{1}$ is a constant.
In this way, we have

\begin{equation}
\rm \dot{P}_{c}=4\alpha + 8p_{a} P_{c}, \qquad
P_{c}=\frac{P_{0}}{8p_{a}}e^{8p_{a}t} - \frac{\alpha}{2p_{a}},
\end{equation}
with $P_{0}$ an integration constant. So, by re-introducing this one
 into the last system of differential equations it is
possible to solve them
 $$\rm c(t)=2p_{a}\,t + c_{o}, \qquad \rightarrow C(t)=C_{0}e^{2p_{a}t}$$.
$$\rm
a(t)= a_0 + \frac{b_{0}}{4p_{a}} t +
\frac{P_{0}}{32p_{a}^2}e^{8p_{a}t}, \qquad A(t)=A_{0}
e^{\frac{b_{0}}{4p_{a}}t}\,\,
{exp}\left[2\frac{C_{0}^4}{p_{a}^2}e^{8p_{a}t}\right].
$$
$$\rm
\phi(\tau)=\phi_0 t + \phi_1,\qquad\qquad \sigma(\tau)=\sigma_0 t +
\sigma_1, $$
 where $(\phi_1,\sigma_1)$ are
integration constants, whereas $\rm \phi_0=-\frac{32}{11}p_\phi+
\frac{6}{11}p_\sigma$ and $\rm
\sigma_0=\frac{6}{11}p_\phi-\frac{2}{33}p_\sigma$ and the constant
$b_{0}=\phi_{0}^2 + 18\phi_{0}\sigma_{0} + 512\pi G \rho_{1} -
4P_{a}^2 + 48\sigma_{0}^2$, and $\rm C_{0}$ is an integration
constant. The last solutions were introduced in the Einstein field
equation, and found that they  being satisfied when $\rm P_{0}=64C_{0}^4$.

Therefore,  the radii $\rm A(t)$ and $\rm C(t)$ have the following
behavior

\begin{equation}\rm
A(t)=A_{0} e^{\frac{b_{0}}{4p_{a}}t}\,\,
\text{exp}\left[2\frac{C_{0}^4}{p_{a}^2}e^{8p_{a}t}\right], \qquad C(t)=C_{0}e^{2p_{a}t}.
\label{radiisol}
\end{equation}

In the frame of classical analysis we obtained that two radii of the
Bianchi $VI_{h=-1}$ type
 are equal ($\rm B(t)=C(t)$), presenting a hidden symmetry, which say that the volume function of this cosmological model
 goes as a jet in the x direction, whereas in the (y,z) plane goes as a circle. This hidden symmetry allows us to simplify the analysis in the classical
 context, and as a  consequence, the quantum version will be
 simplified.

\subsection{Flat FRW embedded in this Bianchi type $\rm VI_{h=-1}$}
In the this subsection we explore the FRW case in this classical
scheme, finding the following.

Solving the standard flat FRW cosmological model in this proposals,
we employed the usual metric
\begin{equation}
\rm ds^2 = -N(t)^2 dt^2 + a(t)^2\left[ dr^2 + r^2 d\theta^2 + r^2
sin^2(\theta) d\phi^2\right], \label{metric-frw}
\end{equation}
where $\rm a(t)$ is the scale factor in this model. The equation
(\ref{stress}) gives the energy density $\rho_{FRW}=M_\gamma
a^{-3(\gamma+1)}$, and the lagrangian density is written as
\begin{equation}\rm
\mathcal{L}_{FRW}= 6\frac{a\dot{a^2}}{N} +2\frac{a^3\dot{\phi^2}}{N}
+ 36\frac{a^3\dot{\phi}\dot{\sigma}}{N} +
96\frac{a^3\dot{\sigma^2}}{N} + 16\pi Ga^3 \rho_{FRW} N.
\label{lagra-frw}
\end{equation}
and the corresponding hamiltonian density in the gauge $\rm N=24a^3$
 and $\gamma=1$, we use the transformation $\rm a=e^A$,
 \begin{equation}\rm
\mathcal{H}_{FRW}=-P_{A}^2  - \frac{48}{11}P_{\phi}^2 +
\frac{18}{11}P_{\phi}\,P_{\sigma} - \frac{1}{11}P_{\sigma}^2  - c_1,
 \label{ham-frw}
\end{equation}
with $\rm c_1=384 \pi G M_1$. The Hamilton equations are
\begin{subequations}
\begin{eqnarray}
\dot{A}&=& -2P_{A}, \label{aa}\\
\rm \dot \phi &=& \rm -\frac{96}{11}P_\phi+
\frac{18}{11}P_\sigma, \label{f-phii}\\
\rm \dot \sigma &=& \rm \frac{18}{11}P_\phi-\frac{2}{11}P_\sigma, \label{f-sigmaa}\\
\rm \dot{P}_A&=& \rm  0, \qquad \qquad \qquad \Rightarrow \qquad P_A= p_{A}=constant, \label{paa}\\
\rm \dot{P}_{\phi}&=& \rm 0, \qquad \qquad \qquad \Rightarrow \qquad P_\phi=p_{\phi}=constant, \label{pphii}\\
\rm \dot{P}_{\sigma}&=& \rm 0, \qquad \qquad \qquad \Rightarrow
\qquad P_\sigma=p_{\sigma}=constant.
\end{eqnarray}
\label{psigmaa}
\end{subequations}
the solutions for the moduli fields are
\begin{eqnarray}
\rm \phi&=&\rm \varphi_0 t +\varphi_1, \qquad
{\varphi_0}_{FRW}=-\frac{96}{11}p_\phi+ \frac{18}{11}p_\sigma =3\left(-\frac{32}{11}p_\phi+ \frac{6}{11}p_\sigma \right)=3{\phi_0}_{(VI_{h=-1})},\nonumber\\
\rm \sigma&=& \rm \xi_0 t +\xi_1, \qquad
{\xi_0}_{FRW}=\frac{18}{11}p_\phi-\frac{2}{11}p_\sigma
=3\left(-\frac{2}{33}P_\sigma +\frac{6}{11}P_\phi\right)=3
{\sigma_0}_{(VI_{h=-1})}, \nonumber
\end{eqnarray}
and the scale factor A, solving the equation (\ref{aa}) is
\begin{equation}
\rm A(t)=2p_A t + A_1, \qquad \Rightarrow \qquad a(t)=a_0 e^{2p_A t}
\end{equation}
 where $\rm p_A=\sqrt{\frac{4608 \pi G M_1 +
{\varphi_0^2}_{(FRW)} + 18 {\varphi_0}_{(FRW)} {\xi_0}_{(FRW)} + 48
{\xi_0^2}_{(FRW)}}{12}} $. This solution is similar at the radii C=B
in the Bianchi type VI, studied in the previous case. With this
result, we can infer that the flat FRW cosmological model is
embedded in this anisotropic cosmological Bianchi type model.

\section{Quantum scheme}

In section \ref{CLH}, we see that the Hamiltonian $\mathcal{H}$ is a constraint, and the lapse function $N(t)$ is a non-dynamical degree of freedom. The last result tell us that we can extend our model to quantum scheme. This  means that we have  to solve the equation
 $\mathcal{H}\Psi=0$, where $\Psi$ is the wave function of the universe. The wave function is a functional
  $\Psi(\phi,\sigma,A,B)$, where $(\phi,\sigma,A,B)$ are the coordinates of the superspace.
  The last ideas are the basis of the canonical quantization and the equation $\mathcal{H}\Psi=0$
  is known as the Wheeler-DeWitt (WDW) equation. This equation is a second-order differential equation on superspace,
  that means, we have one differential equation in each hypersurface of the extended space-time (for more details
  see \cite{Kiefer:2007ria,moniz2010quantum}). On the other hand, the WDW equation has factor-ordering ambiguities, and
  the derivatives are the Laplacian in the supermetric $\mathcal{G}_{ij}$ \cite{moniz2010quantum}.
\\

There most remarkable question that has been dealt with the WDW
equation is to find a typical wave function of the universe, this
subject was nicely addressed in \cite{Gibbons:1989ru,Hartle:1983ai},
and related with the problem of how the universe emerged from a big
bang singularity can be read in
\cite{kiefer2000conceptual,fang1987quantum}.  One remarkable feature
of the WDW equation is its similarity with Klein-Gordon equation.
In order to achieve the  WDW equation for this model we
shall replace the generalized momenta $\Pi_{q^{\mu}}\to
i\hbar\partial_{q^{\mu}}$ in the Hamiltonian (\ref{ham}), these
momenta are associated to the scale factors $A$ and $C$, and the
moduli fields $(\phi,\sigma)$. The WDW equation for this model can
be built by doing $A^{2}\frac{\partial^2}{\partial A^{2}}\rightarrow
A^{p+2}\frac{\partial}{\partial A}(A^{-p}\frac{\partial}{\partial
A})= A^2\frac{\partial^2}{\partial A^2} - pA\frac{\partial}{\partial
A}$, where $p$ is a parameter which measures the ambiguity in the
factor ordering. In this way, we obtain

\begin{equation}
\hat{\mathcal{H}}\Psi=\hbar^2A^2\frac{\partial^2\Psi}{\partial A^2} - \hbar^2pA\frac{\partial\Psi}{\partial A} - 2\hbar^2AC
\frac{\partial^2\Psi}{\partial A\partial C} + \frac{16\hbar^2}{11}\frac{\partial^2\Psi}{\partial\phi^2} - \frac{6\hbar^2}{11}
\frac{\partial^\Psi}{\partial\phi\partial\sigma} + \frac{\hbar^2}{33}\frac{\partial^2\Psi}{\partial\sigma^2} + (16C^4 + C_{1})\Psi = 0,
\label{ham-0}
\end{equation}
where the constant $C_{1}$ is associated to $\gamma=1$ in the
Hamiltonian (\ref{ham}). The last partial differential equation
 can be rewritten by using the transformations $A=e^{a}$ and $C=e^{c}$, and this can be read as

\begin{equation}
\hat{\mathcal{H}}\Psi=\hbar^2\frac{\partial^2\Psi}{\partial a^2} - \hbar^2(p + 1)\frac{\partial\Psi}{\partial a}
- 2\hbar^2\frac{\partial^2\Psi}{\partial a\partial c} + \frac{16 \hbar^2}{11}
\frac{\partial^2\Psi}{\partial\phi^2} - \frac{6\hbar^2}{11}\frac{\partial^2\Psi}{\partial\phi\partial\sigma} +
\frac{\hbar^2}{33}\frac{\partial^2\Psi}{\partial\sigma^2} + \left(16e^{4c} + C_1\right)\Psi =0.
\label{ham-1}
\end{equation}
By looking at  the last expression, we can write the wave function $\Psi(a,c,\phi,\sigma)=\Theta(a,c)\Phi(\phi,\sigma)$
and this give us  two equations in a separated way

\begin{subequations}
\begin{align}
&\frac{\partial^2\Theta}{\partial a^2} - (p + 1)\frac{\partial\Theta}{\partial a} - 2\frac{\partial^2\Theta}{\partial a\partial c}
- \frac{1}{\hbar^2}\left(16e^{4c} + C_{1} + \mu^2\right)\Theta = 0,
\label{theta}\\
&-\frac{16}{11}\frac{\partial^2 \Phi}{\partial\phi^2} + \frac{6}{11}\frac{\partial^2\Phi}{\partial\phi\partial\sigma}
- \frac{1}{33}\frac{\partial^2\Phi}{\partial\sigma^2} + \frac{\mu^2}{\hbar^2}\Phi = 0,
\label{phii}
\end{align}
\label{systems}
\end{subequations}
where $\mu^2$ is the separation constant. The partial differential equation \eqref{theta} associated to the variables $A$ and $C$ has a solution by taking the following ansatz

\begin{equation}
\Theta(a,c)=e^{\alpha a}G(c),
\label{answer}
\end{equation}
where $\alpha$ is a real constant. With this in mind, we obtain

\begin{equation}
\frac{dG}{dc} + \frac{1}{2\hbar^2\alpha}\left(16e^{4c} + \alpha_{0}\right)G = 0,
\label{diffG}
\end{equation}
where we have defined the constant $\alpha_{0}=C_{1} + \mu^2 - \hbar^2\alpha^2 + \hbar^2(p+1)\alpha$, and the solution is

\begin{align}
\Theta(A,C)=G_{0}\biggl(\frac{A}{C^{\beta/\alpha}}\biggr)^{\alpha}e^{-\frac{2}{\hbar^2\alpha}C^{4}},
\label{thetsol}
\end{align}
where $\rm \beta=\frac{\alpha_{0}}{2\hbar^2\alpha}$, and we used the
scale factors $\rm A=e^{a}$ and $\rm C=e^{c}$. On the other hand,
the solutions associated to the moduli fields corresponds to the
hyperbolic partial differential equation \eqref{phii}, whose
solution is given by

\begin{equation}\rm
\Phi(\phi,\sigma)=C_{3}\sin\left[\Lambda(C_{1}\phi + C_{2}\sigma)\right] + C_{4}\cos\left[\Lambda(C_{1}\phi + C_{2}\sigma)\right],
\label{solphi}
\end{equation}
where $\{C_{i}\}_{i=1}^4$ are integration constants, and $\rm
\Lambda=\mu\sqrt{-\frac{33}{48C_{1}^2 - 18C_{1}C_{2} + C_{2}^2}}$.
The last expression \eqref{solphi} has two different behaviors,
these behaviors are given by the cases $48C_{1}^2 - 18C_{1}C_{2} +
C_{2}^2 < 0$ and $48C_{1}^2 - 18C_{1}C_{2} + C_{2}^2
> 0$. For the first case, we have that the behavior of the wave
function associated to the moduli fields $(\phi, \sigma)$ is
oscillatory, and this happens when
 $C_{1}\in\left[\left(\frac{9 - \sqrt{33}}{48}\right)C_{2},\left(\frac{9 + \sqrt{33}}{48}\right)C_{2}\right]$, with $C_{2}>0$. On the other hand, if $C_{2}<0$ we have that the quantum solution becomes hyperbolic functions, and this represents the second kind of solution, we summarize the solutions in the Figure 1. 

\begin{figure}[!htbp]
\begin{center}
\includegraphics[width=0.8\textwidth]{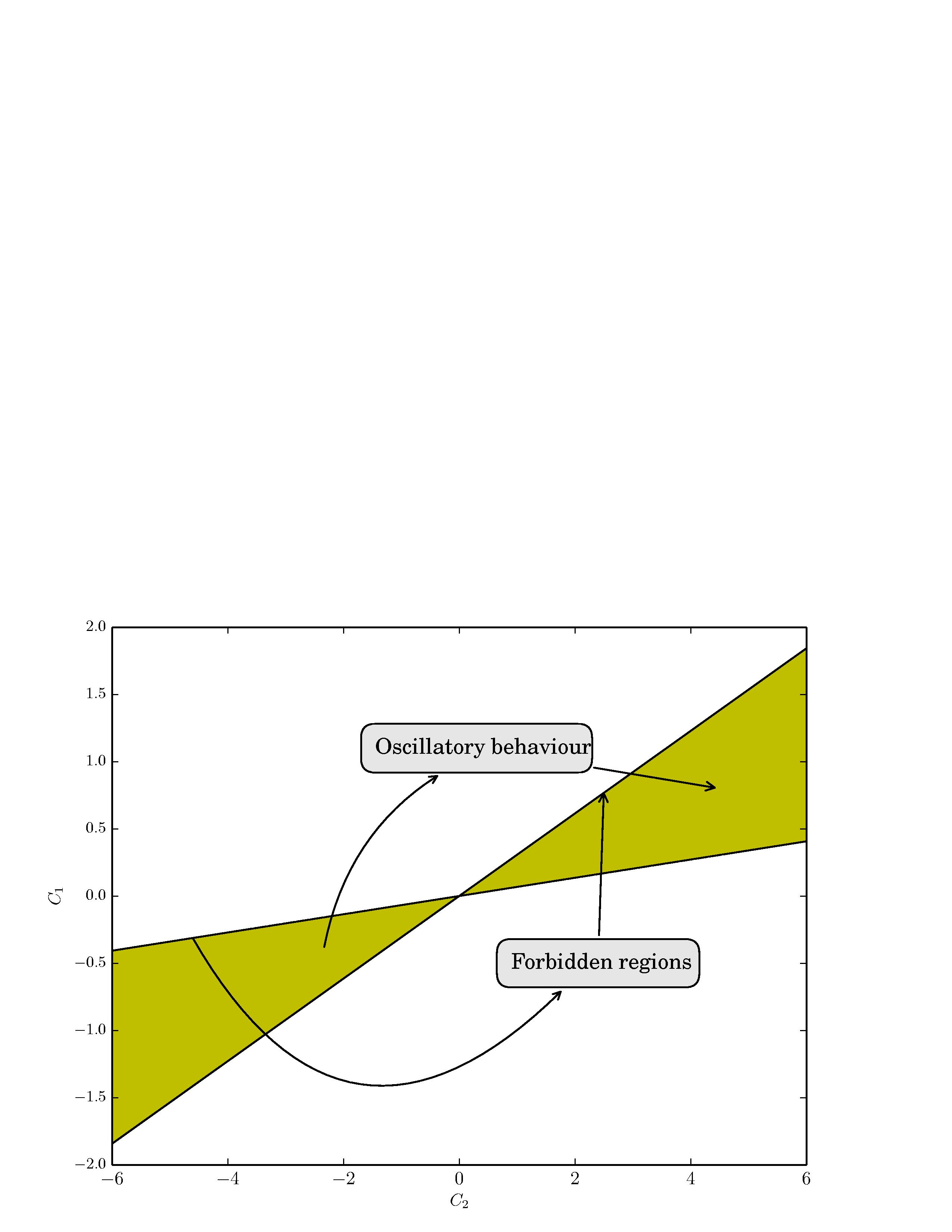}
\caption{Regions where the behavior of the moduli fields are
oscillatory or hyperbolic. The region outside of the oscillatory and
forbidden regions correspond to the hyperbolic behavior.}
\end{center}
\label{beh}
\end{figure}

On the other hand, we see that the solution associated to the equation \eqref{phii} is the same for all Bianchi Class A
cosmological models. It was the main result obtained in \cite{Sesma:2015zua,Socorro:2015zfa}. This can be appreciated by
observing the Hamiltonian operator \eqref{ham-ac}, which can be split as
 $\rm \hat H(a,c,\phi,\sigma)\Psi=\hat H_g(a,c)\Psi + \hat H_m(\phi,\sigma)\Psi=0$,
 where $\rm \hat H_g$ y $\rm \hat H_m$ represents the Hamiltonian for gravitational sector and the moduli fields,
 respectively, and the scale factors have the transformations $A=e^a, C=e^c$.

With the last results, we obtain the general solution to the WDW equation \eqref{ham-1}
 whose wave function $\Phi$ can be built by taking the superposition of the functions
 \eqref{thetsol} and \eqref{solphi}, that is

\begin{equation}
\Psi(A,C,\phi,\sigma)= \mathcal{K}\left(\frac{A}{C^{\beta / \alpha}}\right)^{\alpha}
e^{-\frac{2}{\hbar^2\alpha}C^4}\left(\sin\left[\Lambda(\phi + \sigma)\right] + \cos\left[\Lambda(\phi + \sigma)\right]\right),
\label{genersola}
\end{equation}
where we have taken the constants $G_{0}=1$ and
$\{C_{i}\}_{i=1}^{4}=1$ for simplicity, and $\mathcal{K}$ is a
constant the allows us to normalize the wave function. The
normalization constant $\mathcal{K}$ can be obtained by demanding
the condition

\begin{equation}
\int \Psi(A,C,\phi,\sigma)\Psi^{*}(A,C,\phi,\sigma) = 1.
\label{cond}
\end{equation}

\section{Bohm's formalism }

So far, we have solved the WDW equation \eqref{ham-1}, and we have
found that the wave function $\Psi$ can be split as the product of
two wave functions $\Theta$ and $\Phi$, which are given by
(\ref{thetsol},\ref{solphi}). On the other hand, we know that WKB
approximation is very important in quantum mechanics and therefore,
 we shall introduce an ansatz for the wave function $\Psi$ to
take the  form

\begin{equation}
\Theta(\ell^\mu) = W(\ell^\mu) e^{- \frac{S(\ell^\mu)}{\hbar}},
\label{ans}
\end{equation}
where $W(\ell^{\mu})$ is an amplitude which varies slowly and
$S(\ell^{\mu})$ is the phase whose variation is faster that the
amplitude, this allows us to obtain \textit{eikonal}-like equations.
The term $\ell^{\mu}$ is the dynamical variables of the
minisuperspace which are $\ell^{\mu}=a, c$. This formalism is known
as the Bohm's formalism too
\cite{licata2014quantum,moniz2010quantum,socorro2012quintom,Bohm:1951xw}.

So, the equation \eqref{theta} is transformed under the expression
\eqref{ans} into
\begin{align}
&\hbar^2\left[\frac{\partial^2 W}{\partial a^2}  - (p + 1)\frac{\partial W}{\partial a} - 2 \frac{\partial^2 W}{\partial a\partial c}\right]
+ \hbar\left[-2 \frac{\partial W}{\partial a}\frac{\partial S}{\partial a} - W\frac{\partial^2 S}{\partial a^2}
+ (p + 1)W\frac{\partial S}{\partial a} +2\frac{\partial W}{\partial a}\frac{\partial S}{\partial c}
+ 2\frac{\partial W}{\partial c}\frac{\partial S}{\partial a} + 2W\frac{\partial^2S}{\partial a\partial c}\right] \nonumber\\
& + W \left[\left( \frac{\partial S}{\partial a}\right)^2
- 2\left(\frac{\partial S}{\partial a} \right) \left(\frac{\partial S}{\partial c} \right) - (16e^{4c} + C_{1} +\mu^2)\right]= 0,
\label{mod}
\end{align}
The last expression \eqref{mod} can be written as the following set of partial differential equations, as a WKB-like procedure
\begin{subequations}
\label{WDWa}
\begin{eqnarray}
\left(\frac{\partial S}{\partial a}\right)^2 - 2\left(\frac{\partial S}{\partial a} \right) \left(\frac{\partial S}{\partial c} \right)
- (16e^{4c} + C_{1} + \mu^2) &=& 0, \label{hj}\\
-2 \frac{\partial W}{\partial a}\frac{\partial S}{\partial a} - W\frac{\partial^2 S}{\partial a^2}
+ (p + 1)W\frac{\partial S}{\partial a} +2\frac{\partial W}{\partial a}\frac{\partial S}{\partial c}
+2 \frac{\partial W}{\partial c}\frac{\partial S}{\partial a} + 2W\frac{\partial^2 S}{\partial a\partial c} &=& 0, \label{wdwho}\\
\frac{\partial^2 W}{\partial a^2} - (p + 1)\frac{\partial W}{\partial a} - 2 \frac{\partial^2 W}{\partial a \partial c} & = & 0.
\label{cons}
\end{eqnarray}
\end{subequations}
This set of equation are solved in a different way with the previous
works, because the constraint equation was the equation (\ref{cons})
in the past, however now is solved in first instance. In this set of
equations, the first partial differential equation is known as the
Hamilton-Jacobi equation for the gravitational
 field, with the equation (\ref{cons}) we obtain the W function,  and the the expression \eqref{wdwho} is the constraint equation.
 From the last system of partial
 differential equation \eqref{WDWa} , we observe that the equation \eqref{cons} has a solution given by
\textbf{}
\begin{equation}
\rm W(a,c)= \frac{1}{S_0^2}\left[4 S_4 e^{4c}+(2S_4 a - S_5)S_0^2
+S_4(C_1 + \mu^2 + S_0^2) c \right] e^{-\frac{1+p}{2}c}
\label{solcons}
\end{equation}
and the solutions for the function S are
\begin{equation}
\rm S=S_0 a + S_2 c +S_3 e^{4c}, \qquad S_3=-\frac{2}{S_0}, \quad
S_2= \frac{S_0^2-C_1-\mu^2}{2S_0}, \label{s}
\end{equation}
where $\rm S_0, S_4$ and $\rm S_5$ are integration constants. When
we introduce these results into the  equation (\ref{theta}), with
$\rm \Theta(\ell^\mu) = W(\ell^\mu) e^{-
\frac{S(\ell^\mu)}{\hbar}}$, this partial differential equation is
satisfy identically.

\section{Final Remarks}

In this work we have developed the anisotropic Bianchi $VI_{h=-1}$ model from a higher-dimensional theory of gravity,
 our analysis cover the classical and quantum aspects. In the frame of classical analysis we  obtained that two radii of the
  Bianchi $VI_{h=-1}$ type
 are proportional, we choose the equality ($\rm B(t)=C(t)$) presenting a hidden symmetry, which say that the volume function of this cosmological model
 goes as a jet in x direction, whereas in the (y,z) plane goes as a circle. This hidden symmetry allows us to simplify the analysis in the classical
 context, and as a  consequence the quantum version was simplified too.
 The last result we can compared it with the jet emission that occurs in some stars, in this sense we explore in this context the flat FRW
 cosmological model, finding that the scale factor goes to the B radii corresponding to Bianchi type VI cosmological model,
we can infer that the flat FRW cosmological model is embedded in
this anisotropic cosmological Bianchi type model. Concerning to the
quantum scheme,
  we can observe that this anisotropic model is completely integrable with no need to use  numerical methods.
  Some results have been obtained by considering just the gravitational variables in \cite{Socorro:2010cr}.
   On the other hand, we obtain that the solutions in the moduli fields are the same for all Bianchi
   Class A cosmological models \eqref{solphi}, the last conclusion is possible because the Hamiltonian
   operator in \eqref{ham} can be written in separated way as $\hat H(A,C,\phi,\sigma)\Psi=\hat H_g(A,C)\Psi + \hat H_m(\phi,\sigma)\Psi=0$,
    where $\hat H_g$ y $\hat H_m$ are the Hamiltonian for the gravitational sector and the moduli fields, respectively.
    The full wave function given by $\Psi =\Phi(\phi,\sigma) \Theta(A,C)$ is a superposition of the gravitational variables
    and the moduli fields. There are some recent research related on this line \cite{Alves-Junior:2016vpw}, they built a
      wave packet for an arbitrary anisotropic background. However, one of the main problem in quantum cosmology is how
     to built a  wave packet that allows to determine the possible states of the classical universe
\cite{bojowald2011quantum,halliwell1990introductory,DeWitt:1967yk,Lemos:1995bs,Blyth:1975is,Farajollahi:2010ni,Letelier:2010un,Vakili:2012nh,Chodos:1979vk}.

 \acknowledgments{
\noindent This work was partially supported by CONACYT  167335,
179881, 237351  grants. PROMEP grants UGTO-CA-3 and UAM-I-43. This work is
part of the collaboration within the Instituto Avanzado de
Cosmolog\'{\i}a and Red PROMEP: Gravitation and Mathematical Physics
under project {\it Quantum aspects of gravity in cosmological
models, phenomenology and geometry of space-time}. Many calculations
where done by Symbolic Program REDUCE 3.8.}.

\appendix \label{appe}
\section{The Explicit Einstein's equations}

The Einstein's equations and the equation of motions are given by the expressions \eqref{EKG}.
In detail they are

\begin{subequations}
\label{systeq}
\begin{eqnarray}
G^0_0:&& \frac{\dot A}{NA}\frac{\dot B}{NB} +\frac{\dot{A}}{NA}\frac{\dot C}{NC} + \frac{\dot B}{NB}\frac{\dot C}{NC}
- \left(\frac{\dot\phi}{N}\right)^2 - 18 \frac{\dot\phi}{N}\frac{\dot\sigma}{N} - 48\left(\frac{\dot\sigma}{N}\right)^2 - 8\pi G \rho
- \frac{1}{A^2}=0, \label{00} \\
G^0_1&=&-G^1_0: \frac{\dot B}{NB} - \frac{\dot C}{NC}=0, \label{01}\\
G^1_1:&& \frac{\ddot B}{N^2B} + \frac{\ddot C}{N^2C} + \frac{\dot B}{NB} \frac{\dot C}{NC} - \frac{\dot B}{NB}\frac{\dot N}{N^2}
- \frac{\dot C}{NC}\frac{\dot N}{N^2} + \left(\frac{\dot\phi}{N}\right)^2 + 18\frac{\dot\phi}{N}\frac{\dot\sigma}{N}
+ 48\left(\frac{\dot\sigma}{N}\right)^2 \nonumber\\
&& + 8\pi G p + \frac{1}{A^2}=0,\label{11}\\
G^2_2:&& \frac{\ddot A}{N^2A} + \frac{\ddot C}{N^2C} + \frac{\dot A}{NA}\frac{\dot C}{NC} - \frac{\dot A}{NA}\frac{\dot N}{N^2}
- \frac{\dot C}{NC}\frac{\dot N}{N^2} + \left(\frac{\dot\phi}{N}\right)^2 + 18\frac{\dot\phi}{N}\frac{\dot\sigma}{N}
+ 48\left(\frac{\dot\sigma}{N}\right)^2\nonumber\\
&& + 8\pi G p - \frac{1}{A^2}=0,\label{22}\\
G^3_3:&& \frac{\ddot A}{N^2A} + \frac{\ddot B}{N^2B} + \frac{\dot A}{NA}\frac{\dot B}{NB} - \frac{\dot A}{NA}\frac{\dot N}{N^2}
- \frac{\dot B}{NB}\frac{\dot N}{N^2} + \left(\frac{\dot\phi}{N}\right)^2 + 18\frac{\dot\phi}{N}\frac{\dot\sigma}{N}
+ 48\left(\frac{\dot\sigma}{N}\right)^2\nonumber\\
&& + 8\pi G p - \frac{1}{A^2}=0,\label{33}\\
\Box \phi=0:&& \left(\frac{\dot A}{NA} + \frac{\dot B}{NB} + \frac{\dot A}{NA} + \frac{\dot C}{NC}\right)\frac{\dot\phi}{N}
+ \frac{\ddot\phi}{N^2} - \frac{\dot\phi}{N}\frac{\dot N}{N^2}=0, \label{dalaphi0}\\
\Box \sigma=0:&& \left(\frac{\dot A}{NA} + \frac{\dot B}{NB} + \frac{\dot A}{NA} + \frac{\dot C}{NC}\right)\frac{\dot\sigma}{N}
+ \frac{\ddot\sigma}{N^2} - \frac{\dot\sigma}{N}\frac{\dot N}{N^2}=0. \label{dalasigma0}
\end{eqnarray}
\end{subequations}
The expressions that involve $\dot{q}$ means derivative with respect to the cosmic time, and the relation between the cosmic and proper times is given by
\begin{equation}
d\tau=N(t)\,dt,
\end{equation}
With the last consideration we see that is possible to rewrite the last system of differential equations \eqref{systeq} as follows
\begin{equation}
\dot{\phi} = N\phi', \qquad \ddot{\phi}= N^2\phi'' + \dot{N}\phi',
\label{correspondence}
\end{equation}
just to to mention the transformation of the scalar field. Under
this change, we see that the system \eqref{systeq} can be
transformed into
\begin{subequations}
\label{redsyst}
\begin{eqnarray}
\label{00M} G^0_0:&& 2\frac{ A'C'}{AC} + \biggl(\frac{C'}{C}\biggr)^2 - \phi'^2 - 18\phi'\sigma' - 48\sigma'^2 - 8\pi G\rho - \frac{1}{A^2} = 0,\\
G^1_1:&& 2\frac{C''}{C} + C'^2 + \phi'^2 + 18\phi'\sigma' + 48\sigma'^2 + 8\pi G p + \frac{1}{A^2}=0,\label{11M}\\
G^2_2 = G^3_3:&& \frac{A''}{A} + \frac{C''}{C} + A'C' + \phi'^2 + 18\phi'\sigma' + 48\sigma'^2 + 8\pi G p - \frac{1}{A^2}=0,\label{22M}\\
\label{dalaM} \Box\phi=0: && 2\,\left(\frac{A'}{A} + \frac{C'}{C}\right)\phi' + \phi''=0, \\
\label{dalasM} \Box\sigma=0: && 2\,\left(\frac{A'}{A} + \frac{C'}{C}\right)\sigma' + \sigma''=0,
\end{eqnarray}
\end{subequations}
where we have use that the scale factors $B(t)$ and $C(t)$ are equal.
This result can be obtained from the expression \eqref{01},
whose differential equation can be transformed into the proper time as
\begin{equation}
\frac{B'}{B} - \frac{C'}{C} = 0 \implies \frac{d}{dt}\ln B - \frac{d}{dt}\ln C = 0 \implies \frac{d}{dt}\ln\left(\frac{B}{C}\right) = 0
 \implies \ln\left(\frac{B}{C}\right) = 1.
\end{equation}
In the last expression we have fixed that the integration constant equal to one.
\bibliographystyle{apsrev}
\bibliography{BianchiVIbiblio}

\end{document}